\documentclass[
final,
5p,
times,
twocolumn,
amssymb
]{elsarticle}

\usepackage{bm}
\usepackage{dcolumn}
\usepackage{graphicx}
\usepackage[utf8]{inputenc}
\usepackage[english]{babel}
\usepackage{amssymb}
\usepackage{amsmath}
\usepackage{amssymb}
\usepackage{mathtools, cuted}

\newenvironment{equationc}{\begin{equation}}{,\end{equation}\ignorespacesafterend}
\newenvironment{equationp}{\begin{equation}}{.\end{equation}\ignorespacesafterend}

\begin{document}

\begin{frontmatter}

\title{Searching for Solitons in Heavy-Ion Reactions near the Fermi Energy}


\author{ T. Depastas$^{a,*}$, A. Bonasera$^{a,b}$ and J.B. Natowitz$^{a}$}


\address{ $^{a}$ Cyclotron Institute, Texas A\&M University,
                     College Station, Texas, USA }
\address{ $^{b}$ Laboratori Nazionali del Sud, INFN, Catania 95123, Italy }

\address{ $^{*}$ Corresponding author. Email:  tdepastas@tamu.edu}

\begin{abstract}
Solitons are special shape-conserving hydrodynamical solutions that appear in many areas of physics. Here, we explore the existence of such solutions in microscopic descriptions of the heavy ion reaction $^{12}$C + $^{28}$Si $\rightarrow$ $^{12}$C$^*$ + 7$\alpha$ in the range $E/A=10-65$ MeV/u. After recognizing the centrality of the collision and time-reversibility as fundamental requirements for the presence of solitonic $^{12}$C, we utilize the Hybrid $\alpha$-Cluster model for our analysis with a novel methodology. Our results suggest soliton production for $E/A=25-45$ MeV/u in the forward direction, with a total cross section being at least in the order of a few $\mu$b. This, apart from being encouraging for possible experimental studies, maybe connected to possible toroid states of $^{28}$Si at low angular momenta.
\end{abstract}


\end{frontmatter}

\begin{figure}[!ht]                                        
\centering
\includegraphics[height=6.5 cm]
{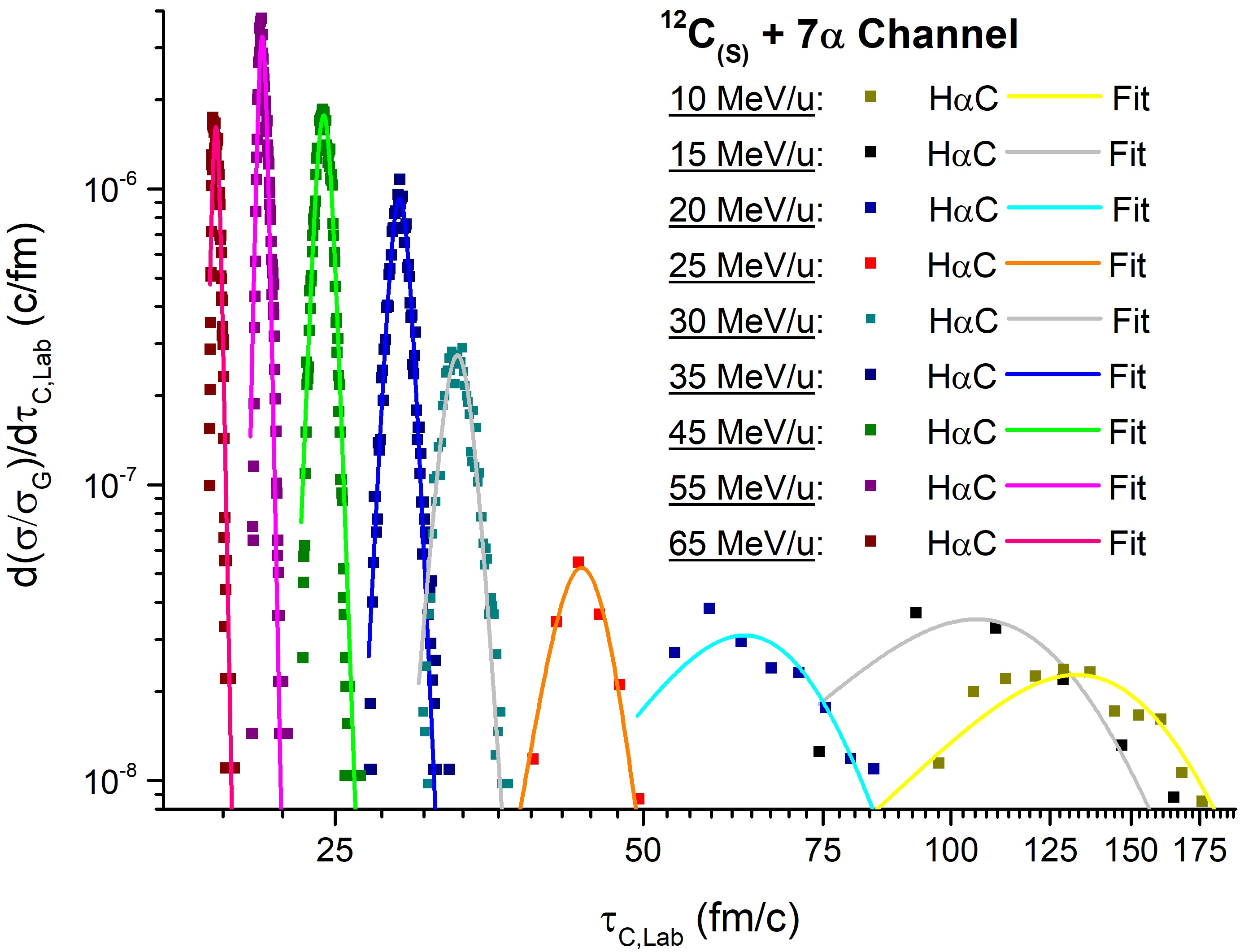} 
\caption{(Color online) $^{12}$C$^*$ appearance probability density as a function of the exit time. The points show results of the H$\alpha$C model and the solid lines correspond to the soliton fits (Eq. \ref{e4}), with different colors for each beam energy, according to the key.}
\label{Fig1}
\end{figure}

\begin{figure}[!ht]                                        
\centering
\includegraphics[height=6.5 cm]
{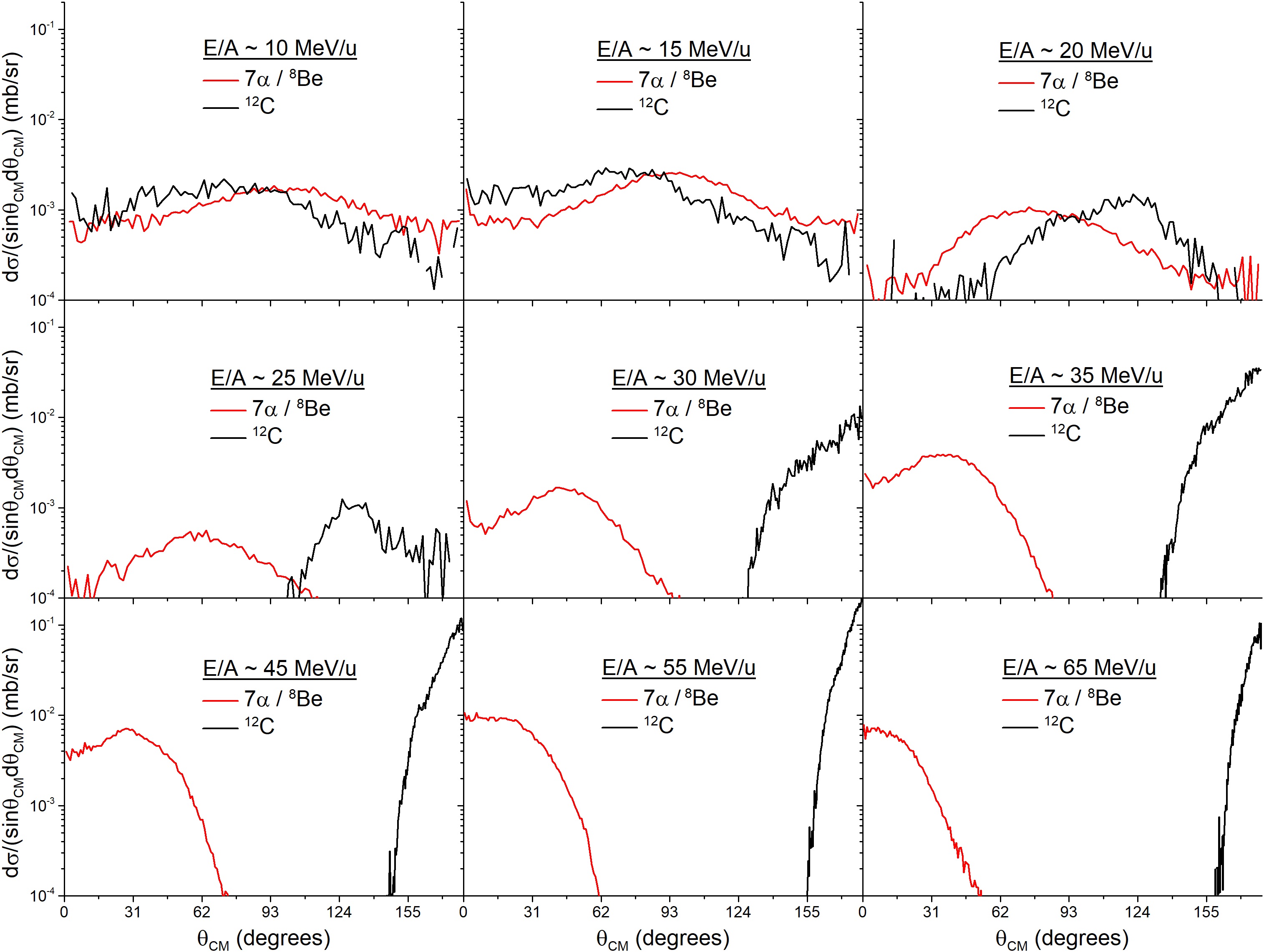} 
\caption{(Color online) Angular differential cross section as a function of the fragment angle in the CM frame. The red line corresponds to the 7$\alpha$'s and the black line to the $^{12}$C. Each panel shows the results of a different beam energy, according to the key.}
\label{Fig2}
\end{figure}

\begin{figure}[!ht]                                        
\centering
\includegraphics[height=6.5 cm]
{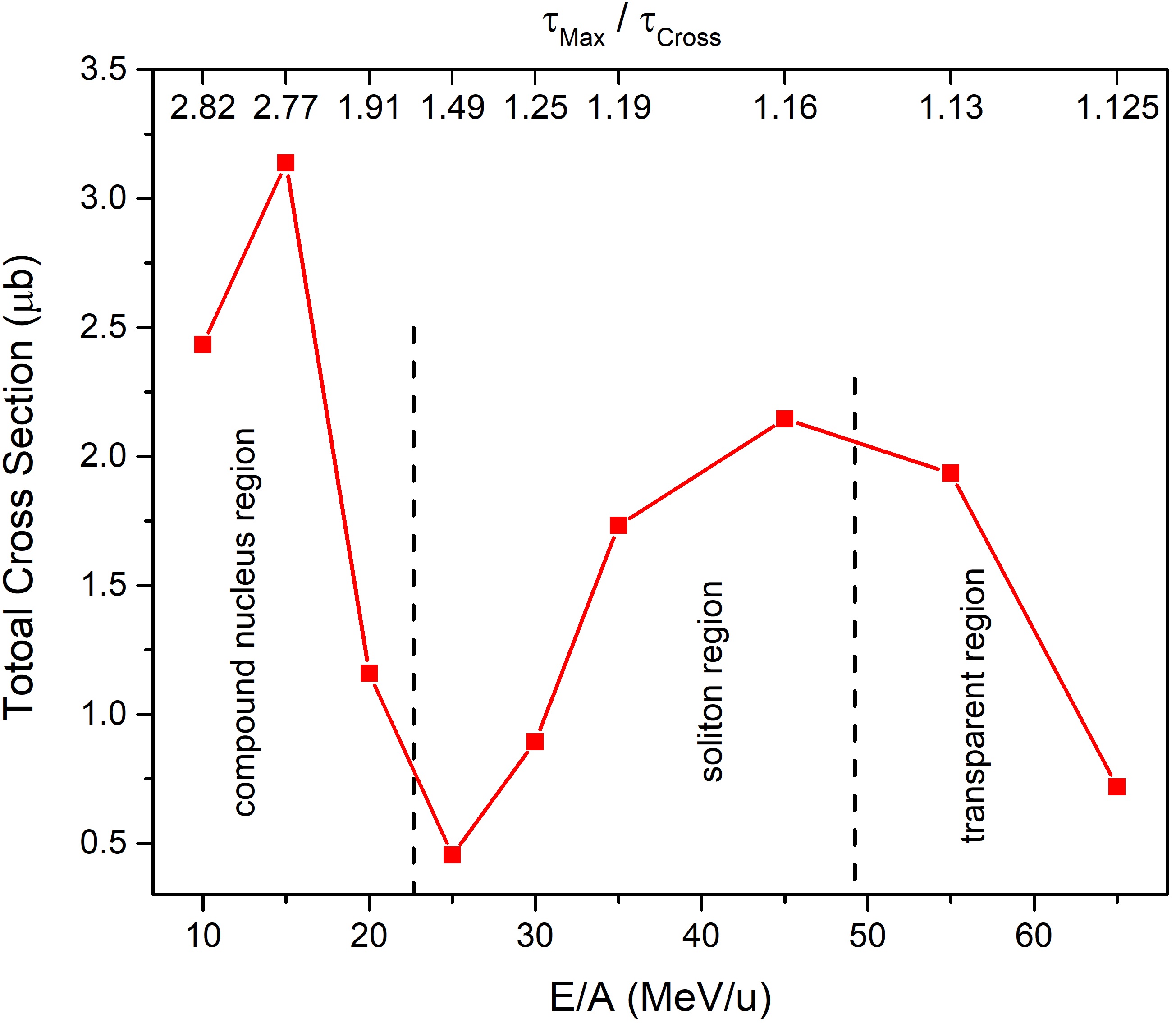} 
\caption{(Color online) Total $^{12}$C$^*$ + 7$\alpha$ cross section as a function of the beam energy. The different mechanism regions are noted by dashed lines. On the top axis, we present the values of $\tau_{Max}/\tau_{cross}$ ratio for each beam energy.}
\label{Fig3}
\end{figure}

\indent Solitary waves or solitons, are special hydrodynamic solutions with very peculiar properties. First discussed by John Scott Russell in the mid 1800's \cite{RussellSolitons1844} as ``waves of translation", solitons are non-linear wave-packets which travel with a constant speed in dispersive medium conserving their amplitude and density profile. Formally, they come as solutions of the Korteweg-de Vries (KdV) equation \cite{KdV} for shallow water waves and their nature is understood as a balance between the dispersive and non-linear terms of the equation, where the former tends to decrease and the latter increase the wave's amplitude. Since their discovery, solitons have been observed in many areas of physics, from plasmas \cite{Zabusky1965,ZhuSubmitted} to boson condensates as solutions of the Gross-Pitaevskii equation \cite{Gross1961,Pitaevskii1961}.\\
\indent In nuclear physics, the Skyrmion model of the nucleon \cite{Skyrme1962SkyrmionModel} as a stable soliton in a pion field, is one of the first instances in which the concept is applied. Continuing into larger scales, Fowler et al. indicate the presence of an entire nucleus as a soliton in heavy ion reactions near the speed of sound \cite{Fowler1982} and discuss several of its characteristics. The authors meticulously prove that the hydrodynamical equations of motion for the reacting nuclear fluid reduce to the KdV equation and admit solitonic solutions. This is done via an expansion of the enthalpy per nucleon up to the second derivative with respect to density, splitting the components parallel and perpendicular to the beam axis and finally, keeping the first order terms of the velocity and density with respect to the slowly moving lab frame \cite{Fowler1982}. Another noteworthy advancement was later achieved by Iwata et al. \cite{Iwata2019,Iwata2020}. The authors provide a detailed description of the solitonic transfer, i.e., a part of the nucleus that considered a soliton is exchanged in a quasi-elastic channel.\\
\indent With this study, we strive to continue the work of Fowler et al., by searching for solitonic solutions in Molecular Dynamics (MD) models and their possible experimental signatures. The structure of the paper is as follows. First, we discuss the necessary conditions for the presence of solitons in microscopic models. Then, we introduce the Hybrid $\alpha$-Cluster (H$\alpha$C) model \cite{ZhengHac2021} and our analytical methodology. Next, we discuss our results that indicate carbon-12 solitons in the $^{12}$C + $^{28}$Si $\rightarrow$ $^{12}$C$^*$ + 7$\alpha$ reaction. Finally, we present our conclusions.\\
\indent From a microscopic point of view, solitons make a key appearance in the Fermi-Pasta-Ulam-Tsingou (FPUT) problem \cite{Benettin2013}. They consider a numerical experiment of non-linear coupled oscillators in one dimension and surprisingly find that the system is non-ergodic. It was later shown \cite{Zabusky1965} that, the equations of motion for the problem can be reduced to the KdV equation in the continuum limit, whose solitonic solutions with their constant amplitude inhibit the system to visit all the points in phase-space and thermalize at long time-scales. This shows then, that solitons may be present only in non-ergodic systems. In MD models, we may extract hydrodynamic effects by ensemble-averaging the trajectories of events with random initial conditions for each time-step, essentially following the Bogoliubov-Born-Green-Kirkwood-Yvon (BBGKY) hierarchy \cite{BonaseraBoltzmann1994}. In principle, microscopic models such as Classical MD and the non-linear Schr{\"o}dinger equation could admit such particle-like solutions. In our study, additional necessary conditions for the observation of solitons are the centrality and energy of the reaction, as well as the time-reversibility of the system.\\
\indent In a heavy-ion reaction, a nucleus can be considered a soliton, if it passes through the other nucleus almost unscathed, as a transmitted disturbance without any loss of nucleons. In such a case, the nucleon-nucleon potential is understood as the source of dispersion and the beam energy as the origin of non-linearity. There is then a minimum required beam energy such that the dispersion is balanced and a soliton appears. This also favors central collisions, as all of the available energy is given to the translational motion. Furthermore, the minimal interaction between the reactants limits our search of the final fragments in the forward direction, i.e., $\theta_{Lab} \sim 0^{\circ}$, as is also hinted in Ref. \cite{Fowler1982}. A wide angular distribution in contrast, would be a signature of a thermalized system. These requirements lead us to exclude fundamental mean field models, such as the Time-Dependent Hartree-Fock (TDHF) as potential calculation tools. This is due to the fact that the mean field, which lacks short range repulsion, allows fast nucleons to travel back and forth in the common potential well at the touching point and reduce the total collective momentum of the nucleus \cite{Flockard1978}. This in turn, renders the system unable to convert high radial translational energies into excitation and creates a non-reactive low impact parameter window \cite{Bonche1979}.\\
\indent The impact parameter restrictions of the TDHF, can be rectified with the use of a collision term \cite{Bonche1979}. This ingredient though, breaks the time-reversibility of the models and produces entropy. The strong connection between the collision term and the absence of solitons can be understood on different levels. Microscopically, collisions are the driving local mechanism of thermalization and ergodicity for a system. As time increases more phase space states are visited and more degrees of freedom are excited, something that can be inferred from the discussion of the ``Loschmidt's paradox" \cite{Darrigol2021}. This essentially makes nucleon-nucleon collisions an internal decoherence mechanism for the nucleus solitary wave. From the Hydrodynamical stand point, time-irreversibility corresponds to a viscous term that causes energy dissipation, which is an additional amplitude decreasing effect and breaks the balance of non-linearity and dispersion. This argument is also implied by Fowler et al. \cite{Fowler1982}, as the equations of motion of choice are the non-viscous Euler equations. The isentropic condition signifies that collision term inclusive models such as the Boltzmann-Uehling-Uhlenbeck (BUU) \cite{BonaseraBoltzmann1994,BulgacPRC2022,ColonnaPRC2021}, Boltzmann-Nordheim-Vlasov (BNV) \cite{BonaseraBNV}, Quantum MD (QMD) \cite{AichelinQMD}, Antisymmetrized MD (AMD) \cite{OnoAMD} and Constrained MD (CoMD) \cite{BonaseraCoMD}, may not admit soliton solutions. While the Fermionic MD (FMD) \cite{FeldmeierFMD} does not have such a term and may in principle allow solitary solutions, its major issue would be the huge computational time required for calculations with very small cross sections. \\
\indent Following these restrictions, we utilize the Hybrid $\alpha$-Cluster (H$\alpha$C) model \cite{ZhengHac2021}. The framework has successfully been used for studies of fusions below \cite{Depastas2023,Depastas2024EPJ,Depastas2024Plb} and above the Coulomb barrier \cite{ZhengHac2021}. It was also employed for simulation of rotational effects \cite{ZhengHac2021} of the silicon-28 nucleus and a dynamical description of the $^{28}$Si + $^{12}$C $\rightarrow$ ... + 7$\alpha$ reaction at 35 MeV/u, where the high 7$\alpha$ excitation energy tail of the cross section is well reproduced at the clusterization dominant region of the data \cite{Depastas7aPRC2025}. The H$\alpha$C model is comprised of $\alpha$ particles as fundamental degrees of freedom that interact via a microscopic pairwise potential consisting of three terms. The Coulomb and Bass \cite{Bass1977} potentials, as well as an effective Fermi energy term, that accounts for the Pauli and Heisenberg correlations of the inner nucleons. The time evolution of the system is then given by Hamiltonian equations of motion, which are of course time-reversible.\\
\indent For our purposes we focus our attention exclusively to the $^{12}$C + $^{28}$Si $\rightarrow$ $^{12}$C$^*$ + 7$\alpha$ channel at 10 MeV/u, 15 MeV/u, 20 MeV/u, 25 MeV/u, 35 MeV/u, 45 MeV/u, 55 MeV/u and 65 MeV/u beam energies and look for $^{12}$C$^*$ solitons, where there is no particle exchange, i.e., all the 7$\alpha$ come originally from silicon. The possibility of carbon to be excited is allowed, which implies decay to 3$\alpha$ at and above the Hoyle state. We stress that this is far from the only channel that may exhibit such solutions. In the $^{28}$Si + 3$\alpha$ channel $^{28}$Si could be the soliton and in the quasi-elastic $^{12}$C$^*$ + $^{28}$Si$^*$ channel where both fragments can also be considered as solitons. This is a particular example of an N-soliton system \cite{Manukure2021}, which by itself possesses several interesting properties, such as spectra sorted by the soliton masses \cite{Landau2012Kinetics}, but is beyond the scope of this paper. The additional requirement, for no particle exchange, renders our results as lower limits to experimental measurements.\\
\indent As we already stated, solitons appear in central collisions, so we simulate reactions with initial relative angular momentum $l=0$. Using randomly rotated initial conditions for the reactant species, an ensemble of $10^6$ events for each energy is generated. Each event evolves according to the equations of motion for $1100$-$2600$ fm/c (according to the beam energy) and the final fragments  are identified via a simple proximity algorithm, similar to Ref. \cite{BonaseraCoMD}. We fix our reference to the lab frame of the $^{12}$C, i.e., where the $^{28}$Si is initially stationary and look at the point of exit of $^{12}$C from the $^{28}$Si within the range of $2R_{Si}=6.98$ fm. There, we define the exit time ($\tau_{C,Lab}$) and characteristic crossing time ($\tau_{cross}$): 
\begin{equationc}
\tau_{C,Lab}=\frac{2R_{Si}}{\gamma\left(v_{C,Lab}\right)v_{C,Lab}}
\label{e1}
\end{equationc}
\begin{equationc}
\tau_{cross}=\frac{2R_{Si}}{\gamma\left(v_{C,0}\right)v_{C,0}}
\label{e2}
\end{equationc}
where $v_{C,0}$ and $v_{C,Lab}$ are initial and final the lab-frame velocities of carbon, respectively and $\gamma\left(v\right)$ is the Lorentz factor. The use of relativistic formulas for the velocities, ensures an accurate description at higher energies, that is, when $v_{C,0} \rightarrow c$ and $\tau_{C,Lab}/\tau_{cross} \rightarrow 0$, as is appropriate for two photon-like events on the Minkowski world-line. The events are then binned according to their $\tau_{C,Lab}$ value and differential cross section per unit of exit time is calculated via the formula (for $l=0$):
\begin{equationc}
\frac{d\sigma}{d\tau_{C,Lab}}=\frac{\pi}{k_{CM}^2}\frac{\delta N}{N \delta \tau_{C,Lab}}
\label{e3}
\end{equationc}
where $k_{CM}$ is the relative wavenumber in the center of mass (CM) frame and $\delta N$ is the number of events out of $N$ events total in the exit time bin of size $\delta \tau_{C,Lab}$. This cross section can be understood as the time evolution of the carbon at fixed spacial point, if we define its appearance probability as $\sigma/\sigma_G$, where $\sigma_G=\pi\left(R_{Si}+R_{C}\right)^2$ is the geometric cross section.\\
\indent The first step in our search for solitons, is attempting to fit the probability density of carbon ($\frac{d\left(\sigma/\sigma_G\right)}{d\tau_{C,Lab}}$) with the standard soliton density profile \cite{Landau2012Kinetics}:
\begin{equationc}
\frac{d\left(\sigma/\sigma_G\right)}{d\tau_{C,Lab}}=\frac{d\left(\sigma_0/\sigma_G\right)}{d\tau_{C,Lab}}\cosh^{-2}\left[\beta\left(z-u\tau_{C,Lab}\right)\right]
\label{e4}
\end{equationc}
with $\frac{d\left(\sigma_0/\sigma_G\right)}{d\tau_{C,Lab}}$, $\beta$ and $u$ being fitting parameters and $z=2R_{Si}$ kept constant. The meaning of each parameter can be seen from examining the structure of Eq. \ref{e4}. Graphically, it represents a peak with maximum at $\tau_{C,Lab}=\tau_{Max}$, where:
\begin{equation}
\tau_{Max}=\frac{z}{u}=\frac{2R_{Si}}{u}
\label{e5}
\end{equation}
and a Full-Width at Half-Maximum (FWHM, $\delta \tau$):
\begin{equationp}
\delta \tau=\ln{\left(\frac{\sqrt{2}+1}{\sqrt{2}-1}\right)}\frac{1}{\beta u }\approx \frac{1.76}{\beta u }
\label{e6}
\end{equationp}
\indent Additionally, the total velocity of the soliton ($v_s$) is given by \cite{Landau2012Kinetics}:
\begin{equationc}
v_s=u+\frac{z}{3}\frac{d\left(\sigma_0/\sigma_G\right)}{d\tau_{C,Lab}}=u+\frac{2R_{Si}}{3}\frac{d\left(\sigma_0/\sigma_G\right)}{d\tau_{C,Lab}}\gtrsim c_s
\label{e7}
\end{equationc}
which as shown in Ref. \cite{Fowler1982}, is proportional and slightly larger than the speed of sound in nuclear matter ($c_s$). We then observe that the parameters $\beta$ and $u$ are inverse-proportionally connected to the soliton width and have units of space and velocity, respectively. The width $\delta \tau$ signifies the time required for the soliton constituents to spread. The parameter $u$ determines mostly the velocity $v_s$, since $ u \gg 2R_{Si}\frac{d\left(\sigma_0/\sigma_G\right)}{d\tau_{C,Lab}}$. Furthermore, $u$ determines the $\tau_{Max}$, which corresponds to the time that the carbon disturbance need to bypass the silicon. For a soliton to exist and the balance between dispersive and non-linear ``forces" to hold, it must be true that, $\tau_{Max} + \frac{\delta\tau}{2} \sim \tau_{cross}$ \cite{Landau2012Kinetics}. The $\frac{d\left(\sigma_0/\sigma_G\right)}{d\tau_{C,Lab}}$ parameter on the other hand, is the maximum of the probability density and contributes to the total soliton cross section.\\
\indent The results of our fitting solution for each beam energy (different colors according to the key) are summarized in Fig. \ref{Fig1}, with the probability density as a function of exit time shown by points and the fits (Eq. \ref{e4}) by solid lines. As the beam energy increases the peaks become narrower and their centroids move towards shorter times. We note that below 25 MeV/u, the soliton fits are not adequate, which implies non-solitonic $^{12}$C$^*$ in the exit channel, while there is a noticeable gap in the peaks between 20 MeV/u and 25 MeV/u. This could correspond to a change in mechanism. Additionally, for energies higher than 45 MeV/u, the peaks tend to the $\delta$-function limit. This implies that the $^{12}$C$^*$ passes through $^{28}$Si very quickly, with little interaction.\\
\indent In order to further explore our previous observations from the fits, we calculate the angular distributions of the $^{12}$C$^*$ and the 7$\alpha$ from the momenta directions in the CM frame. In the CM frame within the H$\alpha$C model, the silicon is initially on the negative side on the beam ($\hat{z}$-) axis and travels towards the positive side, opposite of carbon. The forward direction ($\theta_{Lab} \sim 0^{\circ}$) for the $^{12}$C$^*$ is at $\theta_{CM} \sim 180^{\circ}$ and for the 7$\alpha$ at $\theta_{CM} \sim 0^{\circ}$. We obtain the angular cross sections $\frac{d\sigma}{\sin\theta_{CM}d\theta_{CM}}$ a binning process similar to Eq. \ref{e3} and plot them in Fig. \ref{Fig2}. There, black lines correspond to the carbon-12 and red lines to the 7$\alpha$, while each panel shows the curves for a different beam energy.\\
\indent For $E/A < 25$ MeV/u, the distributions are wide, with the lowest energies being almost isotropic. This may correspond to a dominance of a quasi-thermalized compound nucleus mechanism, that evaporates $\alpha$'s and carbons almost statistically. In the $25$ MeV/u $\le E/A \le 45$ MeV/u region, the angular distributions of each species can be distinguished and carbon-12 appears in the forward direction. This could potentially define the soliton region. Interestingly, this is somewhat above the $10-20$ MeV/u region, which agrees with the predictions of Fowler et al. \cite{Fowler1982}. One could additionally calculate the Mach number \cite{Landau2013Fluids}, from the peak of the 7$\alpha$ distribution in the lab frame and extract the speed of sound of nuclear matter \cite{Bonasera1988Scaling}. For the purposes of this paper, we only qualitatively notice the proximity of this quantity to the middle of the distribution which places the soliton velocity slightly above the speed of sound, agreeing once more with Ref. \cite{Fowler1982}. A more detailed study of this characteristic is left for a future work. For beam energies at $E/A > 45$ MeV/u, we note that $^{12}$C distributions are almost exclusive in the forward direction. This might be a signature of the model's transparency. As is discussed in Ref. \cite{ZhengHac2021}, the $\alpha$-$\alpha$ potential for the H$\alpha$C model is repulsive at short distances, such that the finite nuclei are not collapsing. This limits though, the amount of translational energy that can be converted to excitation, in a way, similar to, but much less prominent than a TDHF calculation \cite{ZhengHac2021}. This forces an upper limit to the beam energies that ``real" solitons, which interact and are transmitted through the target, can be produced.\\
\indent To provide a stimulus for a possible experimental study, we extract the total cross section for the $l=0$, $^{12}$C$^*$ + 7$\alpha$ no-particle exchange channel, via integration of Fig.'s \ref{Fig1} points. We plot this as a function of the beam energy in Fig. \ref{Fig3}.\\
\indent Our results show that the total cross section rests in the order of a few $\mu$b, which to our knowledge is measurable by current instrumentation. We stress again that our restrictive channel choice drives us to interpret the curve of Fig. \ref{Fig3} as a lower limit of the corresponding experimentally measured quantity. A possible experiment could then be performed by detection of the 7$\alpha$'s in coincidence with the $^{12}$C, in the forward and backward directions, respectively depending of the choice of projectile and target. The detection of $^{12}$C close to $\theta_{Lab} \sim 0^{\circ}$ can be performed via the use of dedicated spectrometers, such as the VAMOS++ at the GANIL facility \cite{VamosGANIL}, with the Hoyle and $2^+_1$, 4.43 MeV states as additional signatures.\\
\indent We furthermore delineate the three regions of interest with dashed lines and show the values of the $\tau_{Max}/\tau_{cross}$ ratio for each energy. For the lowest energies, the exit time is already 2-3 times longer than the typical crossing time, which agrees with the compound nucleus interpretation and the angular distribution of Fig. \ref{Fig2}. On the contrary, at that high energy limit the exit time tends asymptotically to the crossing time. This is another signature of transparency, as the amount of energy imparted on silicon becomes minimal and the carbon passes through essentially unaffected.\\
\indent The existence of carbon solitary waves in the particular reaction can be connected to the presence of toroidal states in silicon, a contested topic in the current literature \cite{Depastas7aPRC2025,Cao7aPRC,Hannaman7aPRC,HannamanEPJ}. As it is hinted in Ref. \cite{Fowler1982}, the soliton in a central collision maybe able to create a ``hot-spot" of localized energy transfer, which corresponds to a geometric hole in the silicon phase space, i.e., a toroid. To examine that prospect, we apply ``cuts" to the collected $^{12}$C$^*$ + 7$\alpha$ events, based on the carbon exit time, in a similar spirit with Ref. \cite{Stone1997}. For these events, we calculate the 7$\alpha$ excitation energy and create a binned yield spectrum normalized to unity, with Poisson errors. To ensure convergence of the spectra peaks and maximum statistics the cut condition for selecting an event is:
\begin{equationp}
\tau_{Max}+\frac{\delta \tau}{8} \le \tau_{C,Lab}
\label{e8}
\end{equationp}
\indent We also plot the transverse wavenumber (i.e., $K_x$ and $K_y$) probability distributions by binning the wavenumber of each $\alpha$ particle and normalizing to unity.\\
\begin{figure*}[!ht]                                        
\centering
\includegraphics[height=8.0 cm]
{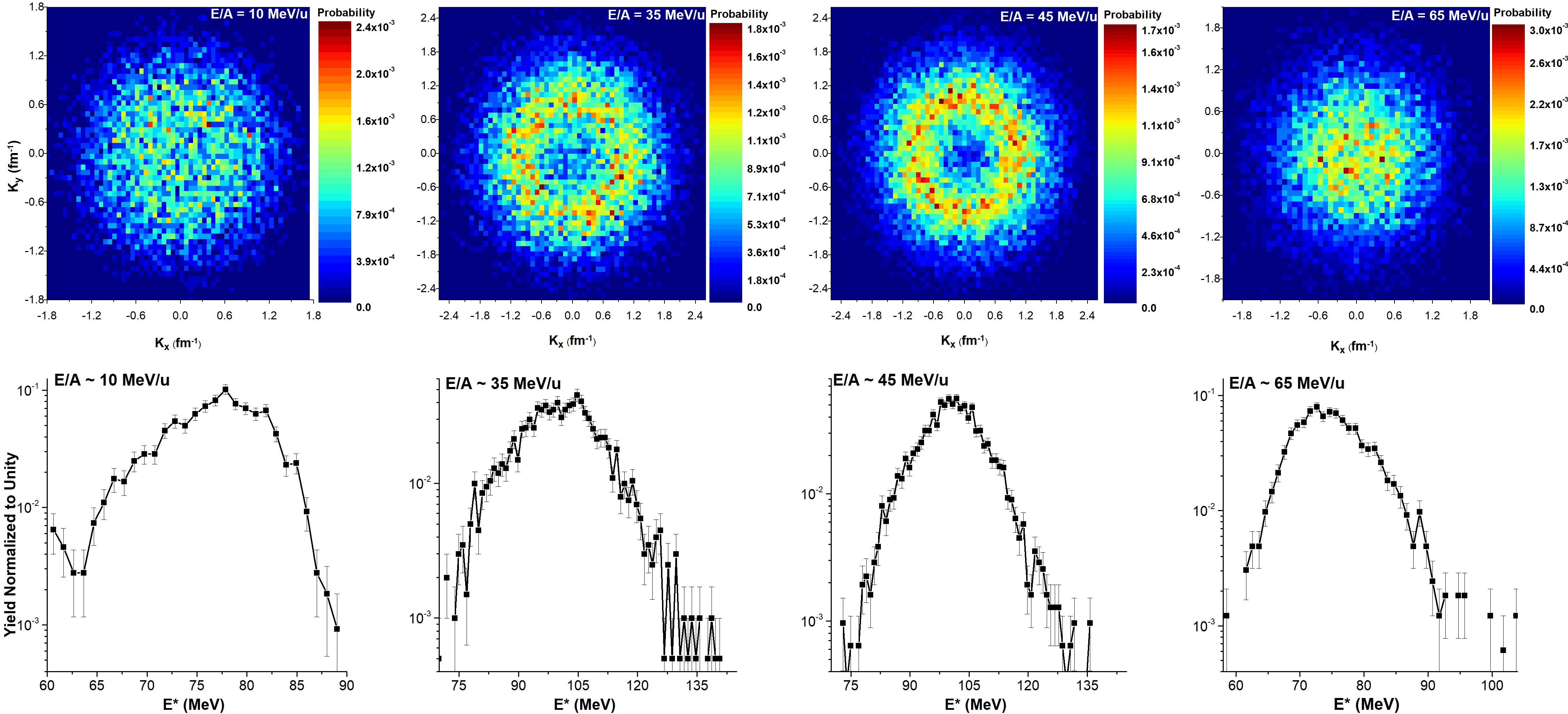} 
\caption{(Color online) Probability distributions of $\alpha$-transverse momenta (top row) and the corresponding 7$\alpha$ excitation energies (bottom) row. Each column corresponds to $E/A$ = 10 MeV/u, 35 MeV/u, 45 MeV/u or 65 MeV/u, from left to right, according to the key.}
\label{Fig4}
\end{figure*}
\indent Our results are collectively depicted in Fig. \ref{Fig4}. The top row shows the transverse momenta distributions as heat-map plots for $E/A$ = 10 MeV/u, 35 MeV/u, 45 MeV/u and 65 MeV/u, while in the bottom row the corresponding 7$\alpha$ excitation energy is plotted. For the lowest beam energy, the $\alpha$-momenta seem isotropic on the transverse plane, the excitation energy is wide and no toroidal state can be seen clearly. These observations are consistent with the compound nucleus description. The high energy extreme also does not show an explicit toroid and the excitation energy is peaked at the same place around $\sim 76$ MeV, but the $\alpha$-momenta are mostly present in the beam direction. This reinforces the idea, that less energy is imparted to the silicon due to transparency. In the soliton region of $E/A$ = 35 MeV/u and 45 MeV/u in contrast, similar structures can be observed. A toroidal-like hot-spot is clearly visible as a hole around the beam axis and the excitation energies in both cases are peaked at $\sim 105$ MeV, essentially the observed peak of the 7$\alpha$ excitation energy distribution for this reaction. Different hydrodynamical instabilities are of course possible which could produce toroidal or other exotic shapes, but we would expect different signatures, beam energies and systems \cite{Bonasera1988Scaling,Natowiz1993,BonaseraTurbulence1996}. Toroidal calculations will be discussed in greater detail in a future paper.\\ 
\indent To conclude, here we study the evidence for possible soliton production in heavy ion reactions, through a microscopic framework. Extending the work of Fowler et al. \cite{Fowler1982}, we understand the balancing of the beam energy and the nuclear interaction as the fundamental mechanism for solitonic production. We furthermore recognize the centrality of the collision, the non-ergodicity and the time-reversibility of the system as necessary requirements for this process.\\
\indent Afterwards, we utilize the Hybrid $\alpha$-Cluster (H$\alpha$C) model to study the possible carbon solitons in the reaction $^{12}$C + $^{28}$Si $\rightarrow$ $^{12}$C$^*$ + 7$\alpha$, with no particle exchange. To that end, we introduce a formalism for the appearance probability density of $^{12}$C$^*$ in the exit channel, as a function of the exit time. We additionally derive the criteria for distinguishing solitonic fragments based on the calculated quantities.\\
\indent Our results suggest the presence of solitons in the region of $E/A \sim 25-45$ MeV/u, where the final fragment is found in the forward direction, resembling the analytical soliton fit and with minimal but non-negligible interaction with the silicon. For the lower energies our evidence points towards a compound nucleus dominance and for the higher energy region, the transparency of our model is prominent. The calculations are encouraging for a possible experimental study of such solitary waves, since our restrictive choice of channel gives a lower limit to the total cross section in the order of a few $\mu$b.\\
As a final note, we explore possible connections between the solitons of carbon and the toroidal states of silicon. This is attempted via a critical selection of events and shows the localized energy transfer from the solitonic carbon to the silicon as geometric hole in phase-space, i.e., a toroid. This should in principle be consistent with the theoretically predicted zero-angular momentum states, but we leave a detailed analysis of this combined effect for a future work.\\ 
\\ACKNOWLEDGMENTS\\
This work work was supported in part by the United States Department of Energy under Grant $\#$DE-FG02-93ER40773.
\bibliographystyle{model1-num-names}
\bibliography{references}{}
\end{document}